\documentclass[aps,prl,twocolumn,showpacs]{revtex4}
\usepackage{amssymb}

\usepackage{graphicx,amsmath}
\usepackage{epstopdf}
\newcommand{\bea}{\begin{eqnarray}}
\newcommand{\eea}{\end{eqnarray}}
\newcommand{\beq}{\begin{equation}}
\newcommand{\eeq}{\end{equation}}
\newcommand{\benu}{\begin{enumerate}}
\newcommand{\enu}{\end{enumerate}}

\begin{document}
%MRN - suggested new title
\title{Kondo Breakdown as a Selective Mott Transition in the Anderson Lattice}
\date{\today}
\author{C. P\'epin}
\affiliation{SPhT, CEA-Saclay, L'Orme des Merisiers, 91191
Gif-sur-Yvette, France\\
%$^2$Materials Science Division, Argonne National Laboratory, Argonne, IL
%60439
}

\begin{abstract}
 We show within the slave boson technique, that the Anderson lattice model exhibits a Kondo
breakdown quantum critical point (KB-QCP) where the hybridization
goes to zero at zero temperature. At this fixed point, the
f-electrons experience as well a selective Mott transition
separating a local-moment phase from a Kondo-screened phase.
 The presence of a
multi-scale QCP in the Anderson lattice in the absence of
magnetism is discussed in the context of heavy fermion compounds. This study is the first evidence for a  selective Mott transition
in the Anderson lattice.
\end{abstract}

\maketitle

Quantum criticality in heavy-electron materials has attracted a
substantial interest recently, mainly triggered by the remarkable
metallic properties of these compounds \cite{stewart}. The
observation of anomalous exponents in both transport (exponents
for the temperature dependence of the resistivity less than two)
and thermodynamics ( the specific heat coefficient doesn't
saturate at low temperatures) contradicts the universal
predictions of the Landau Fermi liquid theories of metals
\cite{review-piers,lohneysen}. It has been suggested early that
deviations from the Landau theory of metals can be explained by
the proximity to a zero temperature phase transition, or QCP.
Close to such a fixed point, the interactions with quantum
critical massless modes substantially shorten the conduction
electron lifetime at the Fermi surface, affecting the observable
properties of the metal. Since heavy fermion compounds are
strongly metallic and magnetic materials, QCPs towards itinerant
antiferromagnetism (AF) have been first studied
\cite{hertz,millis,moriya,rosch}. Although they destabilize the
Fermi liquid, quantum fluctuations in that case are not strong
enough to explain both quasi-linear resistivity and anomalous
thermodynamics properties in 3D. A new class of theories have then
emerged \cite{review-piers,qimiao,ping}. Relying on Doniach's
\cite{doniach} observation that the Kondo effect and the
anti-ferromagnetism are in competition in the Kondo lattice, these
authors suggested that, at the magnetic QCP, another energy scale
vanishes called ``effective Kondo temperature''. This new scale
signals the breakdown of the heavy electron metal. Recently, this
scenario of two energy scales vanishing congruently at the same
point in the phase diagram has been challenged in another
direction. A candidate for the Kondo breakdown quantum critical point (KB-QCP)
 has been found in the
Kondo-Heisenberg lattice model \cite{senthil,schofield,us} as a
fixed point for which the hybridization between impurity electrons
and conduction electrons vanishes. This new fixed point is
intrinsically multi-scale\cite{us}. Two distinct regimes are
distinguished. Below a small energy threshold $E^*$ which depends
on the f- and c- electron band structure (the typical value of
$E^*$
 ranges from 1 mK to 100 mK) thermodynamics and transport are dominated
 by gauge fluctuations \cite{senthil}.
The fluctuations of the order parameter admit a dynamical exponent
$z=2$. Above the scale $E^*$, the fixed point exhibits  marginal
Fermi liquid behavior in $D=3$ (with a dynamical exponent $z=3$).
In this intermediate energy regime the resistivity varies like $T
LogT$. An  important observation is that the Kondo breakdown
relies on the presence of short range antiferromagnetism, which
provides a small bandwidth for the f-electrons. Below a certain
value of the Kondo interaction $J_K$, the dispersion of the
f-electrons de-stabilizes the formation of the heavy metal towards
a spin-liquid phase.  In view of the above observation, there is
no reason why, upon inclusion of magnetism into the model, the  AF
quantum critical point should coincide with the Kondo breakdown.
The mean-field phase diagram rather suggests that the KB-QCP is
generically situated under the AF dome (or as well under any other
kind of instability, like superconductivity). Although it is not
clear at the moment how the low energy regime of the KB-QCP
survives the presence of nearby ordered phases, the intermediate
energy regime, with linear resistivity, is expected to be a robust
feature of the phase diagram.

In this Letter we address the issue of the stability of the KB-QCP
towards charge fluctuations. Our main finding is that, in a slave-boson formulation,
the KB-QCP coincides with a selective Mott transition
 for the f-impurities.   Our study is the first evidence for  a selective Mott transition in
the Anderson lattice. 
 In real heavy fermion compounds, the
number of f-electrons per site is not directly tunable; the
valence of the impurities is allowed to fluctuate.  It is commonly
believed that  compounds
  showing a large
effective mass (of the order of $m^*\simeq 100 \ m_e$ or more) are
in the heavy fermion regime where the charge of the f-impurities
is frozen. It is not clear however, whether the existence of the
KB-QCP is affected by valence fluctuations. To answer this
question we study the Anderson lattice model (where charge is
allowed to fluctuate on the f-impurities) with a small dispersion
of the f-band. 
The situation is similar to the one encountered in the t-J model
of cuprate superconductors at half filling ($\delta=0$ )
\cite{anderson,affleck-marston}, where the spin liquid phase
obtained through a slave-boson formalism is believed to describe
the Mott insulating state of the conduction electrons. In the
Anderson lattice however, the hybridization between the impurity
band and the conduction electron band is driven continuously to
zero at the Mott transition, driving the system through the
KB-QCP.

 The possibility of  a selective Mott
transition in the Anderson lattice has been previously
investigated in the context of single site
DMFT\cite{antoine-gabi}. These authors find that, at zero
temperature,
 an infinitely  small amount of hybridization
destabilizes the Mott transition towards Kondo screening. At
finite temperature, a first order transition terminated by a
critical end-point is obtained. Our results can be reconciled with
those of \cite{antoine-gabi} by noticing that single site DMFT
doesn't account for short-range spin liquid effects. As such, the
hybridization cannot be continuously tuned to zero within this
technique. Our simple analysis suggests  that when lattice effects
are taken into account, the critical end-point of the Mott
transition is tuned to zero for a critical value of the
hybridization. We have also studied the effect of a Coulomb
repulsion $U_{fc}$ between the conduction electrons and the
impurity band. Within our technique $U_{fc}$ doesn't destabilize
the KB-QCP.

We start with the Anderson lattice model with a small dispersion
of the f-band \bea \label{eqn1} H & = & \sum_{\langle i, j \rangle
\sigma} \left ( c^\dagger_{i \sigma} t_{ij} c_{j \sigma} + {\tilde
f}^\dagger_{i \sigma} ( \alpha t_{ij} + E_0 \delta_{ij}) {\tilde
f}_{j \sigma} \right )
 \\
& + & \sum_{i, \sigma} \left ((V {\tilde f}^\dagger_{i \sigma}
c_{i \sigma} + h.c.)   + U {\tilde n}_{f, i}^2 + U_{fc} {\tilde
n}_{f, i} n_{c,i} \right ) \nonumber \ , \eea where $\alpha $ is a small parameter, $\sigma$ is
the spin index belonging to the SU(N) representation, 
$t_{i j}=t$ is the hopping term taken as a
constant, $V$ is the hybridization between the f- and c- bands,
$E_0$ is the energy level of the f-electrons. ${\tilde
n}_{f,i}=\sum_\sigma {\tilde f}^\dagger_{i \sigma} {\tilde f}_{i
\sigma}$ and $n_{c,i}=\sum_\sigma {\tilde c}^\dagger_{i \sigma}
{\tilde c}_{i \sigma}$ are the operators describing the particle
number. We first study (\ref{eqn1}) in the limit of very large  on
site coulomb repulsion U.  In the $U \rightarrow \infty$ limit we
account for the constraint of no double occupancy through a
Coleman\cite{coleman84} boson ${\tilde f}\rightarrow f b^\dagger $
enslaved to a constraint on each site $ \sum_{\sigma} f^\dagger_{i
\sigma} f_{i \sigma} + b^\dagger_i b_i = 1 $\cite{note2}.  Upon this
transformation the effective Lagrangian writes \bea \label{eqn2} L
& = & \sum_{\langle i, j \rangle \sigma} \left ( c^\dagger_{i
\sigma} ( \partial_\tau \ \delta_{ij} + t )  c_{j \sigma} \right . \nonumber \\
& + & \left . f^\dagger_{i \sigma}
(  b_i \alpha t b_j^\dagger + (\partial_\tau + E_0 +
\lambda)\delta_{ij} ) f_{j
\sigma} \right )\nonumber \\
& + &   \sum_i b^\dagger_i \left ( \partial_\tau + \lambda \right
) b_i
-\lambda + \sum_{\langle i, j \rangle} J {\bf S}_{f i } \cdot {\bf S}_{f j}  \\
& + &\sum_{i, \sigma} \left ( ( V  f^\dagger_{i \sigma } b_i c_{i
\sigma } + h.c. )    + U_{fc}
 n_{f, i} n_{c,i} \right ) \nonumber \ , \eea where $ J =2 (\alpha t )^2 /U$,  ${\bf
 S}_{i} = \sum_{\alpha \beta}f^\dagger_\alpha {\bf \sigma}_{\alpha \beta} f_\beta $
 with ${\bf \sigma}$ the
 Pauli matrix. $n_{f, i}= \sum_\alpha
 f^\dagger_\alpha f^\alpha $ is the density operator. The
 constraint has been implemented through a Lagrange multiplier
 $\lambda$.
 The term $J {\bf S}_{f i } \cdot {\bf S}_{f
 j}$ is generated through super-exchange mechanism, as in the t-J model for the cuprate superconductors. 
It is
 insensitive to the slave bosons.
 To proceed we make a static approximation where the phase of the
 slave bosons is frozen.  The super-exchange term is decoupled
 in the uniform-RBV (Resonating Valence Bound) channel, which renormalizes the f-dispersion at
 the Hartree-Fock level,
 $J {\bf S}_{f i } \cdot {\bf S}_{f
 j}\rightarrow f^\dagger_i \beta t f_j $  . $\beta$ is roughly constant
 through the phase diagram\cite{us} and can be approximated by its
 value at the KB-QCP
 \beq
 \beta= \frac{J}{t} = \frac{2 \alpha^2 t}{U} \ . \eeq
  The fc- Coulomb repulsion
 is decoupled using a Hubbard-Stratonovich field ${\vec
 \varphi} $ such that $U_{fc}
 n_{f, i} n_{c,i} \rightarrow {\vec \varphi}_i \cdot c^\dagger_{i
 \alpha } {\vec \sigma}_{\alpha \beta} f_{i \beta}  + \varphi^2/
 U_{fc} $.
 In {\bf k}-space the mean-field equations write
 \bea \label{eqn3}
  & & T  \sum_{k, \sigma, n}  b \alpha \epsilon_k G_{ff}( k, i \omega_n)
  + V T \sum_{k, \sigma, n} G_{fc}( k, i \omega_n) +  b \lambda = 0
  \ , (3.a) \nonumber  \\
 & & T   \sum_{k, \sigma, n}  {\vec \sigma} G_{fc}( k, i \omega_n ) +
{\vec \varphi }/ U_{fc} = 0 \ , (3.b) \nonumber \\
 & & T \sum_{k, \sigma, n}  G_{ff}( k, i \omega_n) + b^2 = N/2 \ , (3.c) \nonumber
 \eea
where $\epsilon_k$ is the dispersion of the c-electrons,
$\epsilon^0_k = \alpha b^2 \epsilon_k  + \beta \epsilon_k + E_0 +
\lambda $ is the dispersion of the f-band\cite{comment1}. $G_{ff}$
and $G_{fc}$ are obtained by diagonalizing the hybridized f- and
c- bands. We first set $U_{fc}=0$ leading to $\varphi=0$ from Eqn(
3.b).
\begin{figure}
\includegraphics[width=2.6in]{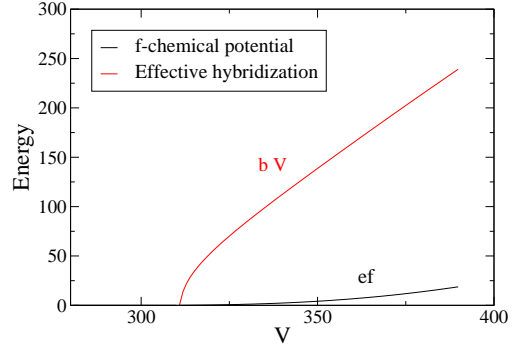}
\caption{ Effective hybridization $V b$ and the f-band chemical potential
$\epsilon_f= E_0 + \lambda$ as a function of V. The electron
bandwidth is $D= 1000$. The chemical potential $\mu = 0$, the
ratio of f- and c- masses is $\alpha =0.1$. $\beta=( 2 \alpha^2 t) /U= 0.01$
and the f-energy level $E_0 = - 500$. The mean field equations are solved for $N=2$. } \label{fig1}
\end{figure}As depicted in figure 1, the set of mean-field equations
admits a QCP where $b \rightarrow 0$ which implies that $n_f
\rightarrow 1$ as the effective hybridization $V b$ goes to zero.
At $n_f=1$, the impurity band is half-filled and the f-electron experience a Mott
transition towards a local state. From (3.a) we
notice that, at the fixed point, $ V T \sum_{k, \sigma, n} G_{fc}(
k, i \omega_n ) = V^2 \rho_0 Log( \beta )  $ with $\rho_0 $ the
density of states of the c-electrons. This leads to the standard
Kondo scale \beq \label{eqn3} \beta_c=  Exp \left [ \frac{E_0}{N
\rho_0 V^2} \right ]\ , \eeq where N is the degeneracy of the
f-electrons. In order to reach continuously the QCP it is crucial
that the spin liquid parameter $\beta$ remains finite through the
phase diagram.

 We turn now to the fluctuations in the quantum critical regime.
In a large N expansion, the fluctuation spectrum is in the same
universality class as the KB-QCP of the Kondo-Heisenberg
model\cite{us}. We recall here the results obtained in this paper.
The KB-QCP exhibits a multi-scale behavior, with $z=2$ dynamical
exponent below $E^*\simeq 0.1 (q*/q)^3 \beta D$, where $q^*=  |
k_F^f -k_F^c|$ is the difference of the Fermi level of the two
species. Above $E^*$,  the physics is dominated by dynamical
exponent $z=3$ and non Fermi liquid behavior is obtained. The
results are summarized in Table 1. Note that the specific heat
coefficient has the same power law dependence as the corrections
to resistivity. This striking property stems form the observation
that in the Mott state, the f-impurities form a reservoir. The
spin liquid description of the Mott state ensures (through gauge
invariance) that, at the QCP, ${\bf v}_f ({\bf r} )=0$ at each
site. In a scattering process with the f-impurities, the momentum
of the light conduction electrons can decay into the reservoir
formed by the heavy f-fermions. Hence, although the boson
propagator admits $z=3$ ( like in the proximity to a ferromagnetic
QCP) the transport lifetime has no extra temperature factor
compared to the electron lifetime.

\vskip0.5 cm
\begin{table}\begin{tabular}{c||c|c|c}
 $ T \gg E^* $ & $ C_v $ & $\Delta \rho (T) $ & $ \chi(T) $\\ \hline
& & & \\
$D=3$ & $- T Log (T) $ & $-T Log(T) $& $ T^{4/3}$ \\ \hline
& & & \\
$D=2 $ & $ T^{2/3} $ & $ T^{2/3}$ & $ -T Log (T) $  \\
\end{tabular} \caption{Transport and thermodynamic exponents in the Maginal Fermi liquid regime around the KB-QCP. 
The exponents are in agreement with those of Ref.\cite{us}.}
\end{table}
\vskip 0.5 cm

In order to study the effect of $U_{fc}$ on the QCP, it is enough
to keep  the component of ${ \vec \varphi} \parallel z$ in (3.b).
 Defining $\Pi_{fc}$ such as $T\sum_{k,n,\sigma} G_{fc} = (b V
 + \varphi) \Pi_{fc}$, and using the change of variables ${\tilde
 \varphi} =  b V + \varphi $, Eqn (3.b) writes
 \beq \label{eqn4} {\tilde \varphi}  \Pi_{fc} + ( {\tilde \varphi} - b V) / U_{fc}
 =0 \ . \eeq   To answer the question of a possible first
 order transition in ${\tilde \varphi}$, we use (3.a) for solving
 for $\Pi_{fc}$, obtaining $ {\tilde \varphi_0} = V - U_{fc} E_0/V
 $. The effective mass for the ${\tilde \varphi}$ field
 $ m_{\tilde \varphi } = E_0/ ( V^2 - U_{fc} E_0) + 1/ U_{fc}$ is
 always positive, thus no first order field driven transition is
 present. However $U_{fc}$ shifts the  QCP, leading to
 \beq \label{eqn5}
 \beta_c = Exp \left [ \frac{ E_0}{N \rho_0 ( V^2 - U_{fc} E_0)}
 \right ] \ . \eeq  Eqn(\ref{eqn5}) interpolates between
 (\ref{eqn3}) for $U_{fc} \ll - V^2/E_0 $ to $ \beta = Exp \left [
 -1/ ( \rho_0 U_{fc} ) \right ] $ for $U_{fc} \gg - V^2/E_0 $.
 Note that  valence transitions
 are known to occur in the mixed valent regime \cite{miyake}.

  To get a deeper insight into the problem, we study the  Mott transition
   as a function of $U$ via
 four Kotliar-Ruckenstein slave bosons\cite{kotliar-ruckenstein}. Since no qualitative changes obtains from the 
inclusion of $U_{fc}$, we proceed with the model at $U_{fc}=0$.
 A set of four creation(annihilation) operators are introduced
 $e_i^\dagger$ ($e_i$), $p_{i \sigma}^\dagger$ ($p_{i \sigma}$), $
 d_i^\dagger$ ($d_i$) which describe respectively zero, one or two
 electrons at the site ``i''. The enlarged Hilbert space is
 restricted by two constraints $\sum_\sigma p_{i \sigma}^\dagger
 p_{i \sigma} + e^\dagger_i e_i + d^\dagger_i d_i =1 $ and $
 f^\dagger_{i \sigma} f_{i \sigma}= p^\dagger_{i \sigma} p_{i
 \sigma} + d^\dagger_i d_i $.  The  Lagrangian (\ref{eqn2}) with $U_{fc}=0$ then takes the form
 \bea \label{eqn6}
 L & = & \sum_{\langle i, j \rangle, \sigma}\left [  c_{i
 \sigma}^\dagger \left (  ( \partial_\tau -\lambda^{(1)} )\ \delta_{ij} + t \right )c_{j \sigma}  \right . \\
  & + & \left . f^\dagger_{i \sigma} (
 z^\dagger_{i \sigma} \alpha  t z_{j \sigma} + \beta t +
 (\partial_\tau + E_0 + \lambda^{(2)}_\sigma ) \delta_{ij} ) f_{j \sigma} \right ]  \nonumber \\
 & + & \sum_i \left
 [ e^\dagger_i ( \partial_\tau + \lambda^{(1)} ) e_i + d^\dagger_i
 ( \partial_\tau + U+  \lambda^{(1)} - \lambda^{(2)}_\sigma ) d_i \right .\nonumber \\
 & + & \left .
 \sum_\sigma p^\dagger_{i \sigma} ( \partial_\tau  + \lambda^{(1)} -
 \lambda^{(2)}_\sigma) p_{i \sigma} \right ] \nonumber \\
 & + & V \sum_{i \sigma} \left ( f^\dagger_{i \sigma} z_{i \sigma} c_{i \sigma} + h.c. \right ) \nonumber \ ,
 \eea where $ z_{i \sigma} = ( 1- d^\dagger_i d_i - p^\dagger_{i
 \sigma} p_{i \sigma} )^{-1/2} ( e^\dagger_i p_i \sigma +
 p^\dagger_{i -\sigma} d_i ) ( 1 - e^\dagger_i e_i - p^\dagger_{i -
 \sigma} p_{i -\sigma} )^{-1/2}$. The form of $z_{i \sigma}$
 ensures that for $U=0$, the average $\langle z^\dagger_{i \sigma}
 z_{i \sigma} \rangle = 1$. The set of mean-field equations is
 obtained by treating the slave -bosons in a static  and uniform
 approximation and by
  differentiating the free energy with respect to
 $\lambda^{(1)}$,$\lambda^{(2)}_\sigma$, $e$,$p_\sigma$,$d$. The result
 is shown in Figure 2. First, let's fix the value of U.
  At $U \geq -E_0$, one reaches a KB-QCP for
 increasing values of $V$. At low $V$ the system is in the
 Mott phase where the impurities are localized, while  above
 $V=V_c$  a finite hybridization sets in,   driving the system to a
 heavy metal fixed point. Alternatively fixing $V$, one obtains a
 line of critical points for
 \beq \label{eqn7}
 U_c = \alpha^2  t \ Exp \left [ \frac{- E_0}{ N \rho_0 V^2} \right ] \ .
 \eeq  For $U \leq U_c$ we are in the Mott phase while for $U \geq
 U_c$ we are in the heavy metal phase. The
 fact that  Mott phase breaks up at high values of $U$
 follows the observation that the spin liquid parameter $J = 2 ( \alpha T)^2 / U $, which
 is necessary to stabilize the  Mott phase, decreases when $U$ increases.
 Following \cite{antoine-gabi}, in the Anderson lattice with no spin liquid ($\beta = 0 $),
 the Kondo
 hybridization always destabilizes the Mott phase towards a screened heavy metal.
 Here, the
higher $V$ is, the lower is the critical $U_c$ at which the KB-QCP
occurs. For $U \ll -E_0$ the Mott transition breaks down. From
DMFT studies\cite{review-antoine}, the $U=-E_0$ line is expected
to be of first order. Note also that for $V=0$ the Mott state
extends to all values of $ U \geq - E_0$, in agreement with
previous studies of the half-filled Hubbard model \cite{lee}.

%\begin{multicols}{2}

\begin{figure}
\includegraphics[width=2.6in]{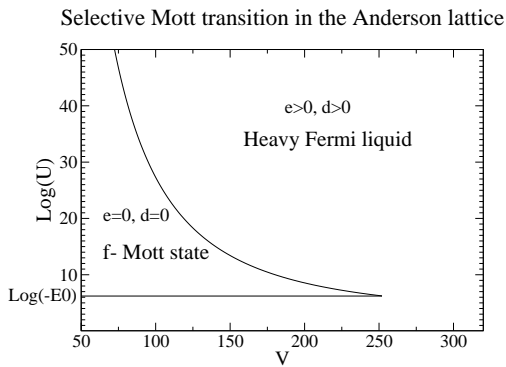}
\caption{ Zero temperature $(Log(U),V)$-phase diagram using four
Kotliar-Ruckenstein bosons. In the intermediate region the f-ipurities 
undergo a Mott transition. The electron bandwidth is $D= 1000$.
The chemical potential is taken to be $\mu=0$, the ratio of f- and
c- masses is $\alpha =0.1$ and the f-energy level $E_0 = - 500$.  Here we take $N=2$. }
\label{fig2}
\end{figure}

We turn now to the discussion of our results in the light of
quantum criticality in heavy fermions. In the standard scenario,
AF fluctuations compete with the formation of the Kondo singlet,
thus preventing the formation of the heavy metal
\cite{review-piers,qimiao}. Although it has been shown that the
one impurity Kondo screening is inhibited by AF fluctuations
\cite{larkin-melnikov,varma}, in the Kondo lattice, however, the
question of whether AF fluctuations are strong enough to destroy
the heavy Fermi liquid remains open. If the standard scenario is
correct, the Kondo breakdown should occur at the point in the
phase diagram where AF fluctuations are maximum, namely at the AF
QCP. Alternatively our study suggests that the Kondo breakdown
occurs at the point where the f-impurities are subject to a
selective Mott transition.  Within our study, a small non
vanishing dispersion of the spinon band  is the necessary and
sufficient condition for the existence of the KB-QCP. The second
scenario thus relies on the presence of a spin -liquid component
of the short AF fluctuations at the Mott transition. The question of the validity of the spin liquid
description of the Mott transition dates from the early days of
high T$_c$ superconductivity with the idea of Resonance Valence
Bond (RVB) around half filling in the Hubbard
model\cite{anderson}. Studies of frustrated magnetism have
concluded that, in the absence of charge fluctuations, spin liquid
phases can be induced by frustration \cite{moessner}. However, no
mixed phase consisting of AF and spin liquid has been found.  In
the presence of charge fluctuations, like around zero doped
cuprate superconductors, it is still unclear whether a short range
RVB state exists or not\cite{lee}. Our study of the Anderson
lattice provides us with a situation where charge fluctuations are
strong  (through coupling to the conduction band), rendering the
occurrence of the spin liquid more favorable.  The presence of the
selective Mott transition in this model is thus a direct test for
the existence of a short range RVB spin liquid, stabilized by
charge fluctuations.

 We thank the hospitality of the KITP(Santa Barbara) for the ``Quantum Phase Transition
 (2005)''
 workshop
 where this work was initiated.
Useful discussions with G. Kotliar, M. Norman, O. Parcollet, I.
Paul and J. Rech are acknowledged.


\begin{thebibliography}{99}
\bibitem{stewart} G. Stewart, Rev. Mod. Phys. {\bf 56}, 755
(1984); {\bf 73}, 797 (2001).
\bibitem{review-piers} P. Coleman {\it et al.}, J. Phys. Cond. Matter
{\bf 13} R723 (2001).
\bibitem{lohneysen} H. v. L\"oneysen {et al.} cond-mat/0606317.
\bibitem{hertz} J. A. Hertz, Phys. Rev. B {\bf 14}, 1165 (1976).
\bibitem{millis} A. J. Millis, Phys. Rev. B {\bf 48}, 7183 (1993).
\bibitem{moriya} T. Moriya and T. Takimoto, J. Phys. Soc.  Japan {\bf 64}, 960 (1995).
\bibitem{rosch} A.  Rosch {\it et al.}, Phys. Rev. Lett. {\bf 79},
159 (1997); A. Rosch {\it ibid} {\bf 82}, 4280 (1999).
\bibitem{qimiao} Q. Si {\it et al.} Nature {\bf 413}, 804 (2001); D.R. Grempel and Q. Si,
Phys. Rev. Lett. {\bf 91}, 026401 (2003).
\bibitem{ping} P. Sun and G. Kotliar, Phys. Rev. Lett. {\bf 91},
037209 (2003).
\bibitem{senthil} T. Senthil {\it et al.}, Phys. Rev. Lett. {\bf 90}, 216403 (2003);
Phys. Rev. B {\bf 69}, 035111 (2004).
\bibitem{doniach} S. Doniach, Physica B, {\bf 91}, 213 (1977).
\bibitem{schofield} P. Coleman, J. B. Marston, A. J. Schofield, Phys. Rev. B {\bf 72}, 245111 (2005).
\bibitem{us} I. Paul, C. P\'epin and M. Norman, Phys. Rev. Lett. {\bf 98}, 026402 (2007).
\bibitem{anderson} P.W. Anderson, Science {\bf 235}, 1196 (1987).
\bibitem{affleck-marston} J. Marston and I Affleck, Phys. Rev. B
{\bf 39}, 11538 (1989).
\bibitem{antoine-gabi} L. de' Medici {\it et al .}  Phys. Rev.
Lett. {\bf 95}, 066402 (2005).
\bibitem{coleman84} P. Coleman, Phys. Rev. B {\bf 29}, 3035
(1984).
\bibitem{note2} Note that the technique with one slave boson treats the low energy part of the model, with no information
about the Hubbard bands.
\bibitem{read} N. Read and D. M. Newns, J. Phys. C {\bf
16}, 3273 (1983); N. Read, J. Phys. C {\bf 18}, 2651 (1985).
\bibitem{comment1} We are considering the uniform condensation of
the slave bosons, occuring when the masses of c- and f-bands are
of the same sign \cite{us}. When masses are of opposite sign the
condensation occurs at finite q.
\bibitem{miyake} A. T. Holmes {\it et al.} Phys. Rev. B {\bf 69}, 024508 (2004); Y. Onishi and
K. Miyake J. Phys. Soc. Japan {\bf 69}, 3955 (2000).
\bibitem{kotliar-ruckenstein} Phys. Rev. Lett. {\bf 57}, 1362
(1986).
\bibitem{review-antoine} A. Georges {\it et al. } Rev. Mod.
Phys.{\bf 68} 13-123 (1996).
\bibitem{lee} see P. A. Lee, N. Nagaosa and X-G. Wen, Rev.
Mod. Phys. {\bf 78}, 17-85 (2006) and Ref. therein.
\bibitem{larkin-melnikov} A. Larkin and M. Mel'nikov, Sov. Phys.
JETP {\bf 34}, 656 (1972).
\bibitem{varma} H. Maebashi, K. Miyake and C . Varma,
Phys.Rev.Lett. {\bf 95},207207 (2005).
%\bibitem{comment2} We
%believe that the spin liquid component  of the AF interaction can
%coexist with the magnetic component, at the Mott transition.
\bibitem{moessner} R. Moessner and S. L Sondhi, Phys. Rev. Lett.
{\bf 86}, 1881 (2001).


\end{thebibliography}
\end{document}